\documentclass[12pt,preprint]{aastex}

\usepackage{graphicx,rotating}
\usepackage{multirow,multicol} 

\newcommand{\RT}{ROTSE3~J160213.1$-$021311.7}
\newcommand{\ntwo}{WFI~J132813.7$-$214237}
\newcommand{\nthree}{WFI~J161953.3+031909}

\def\farcs{\hbox{$.\!\!^{\prime\prime}$}}  				% Fractions of arcseconds
\def\msun{M$_{\odot}$}

\shorttitle{short title}
\shortauthors{Rau et al.}

\begin{document}


\title{The Incidence of Dwarf Novae in Large Area  Transient Searches}


\author{A. Rau\altaffilmark{1}, R. Schwarz\altaffilmark{2}, S.R. Kulkarni\altaffilmark{1}, E.O. Ofek\altaffilmark{1},  M.M. Kasliwal\altaffilmark{1}, C. Brinkworth\altaffilmark{3}, S.B. Cenko\altaffilmark{4}, Y. Lipkin\altaffilmark{5} \& A.M. Soderberg\altaffilmark{1}}
\affil{$^1$ Caltech Optical Observatories, MS 105-24, California Institute of Technology, Pasadena, CA 91125, USA}
\affil{$^2$ Astrophysical Institute Potsdam, An der Sternwarte 16, 14482 Potsdam,  Germany}
\affil{$^3$ Spitzer Science Center, California Institute of Technology, Pasadena, CA 91125, USA}
\affil{$^4$ Space Radiation Laboratory, California Institute of Technology, Pasadena, CA 91125, USA}
\affil{$^5$ School of Physics and Astronomy and Wise Observatory, Tel-Aviv University, Tel-Aviv 69978, Israel}

\email{arne@astro.caltech.edu}

\begin{abstract}

Understanding and quantifying the contribution of known classes of transient and variable sources is an important lesson to be learned from the manifold of pre-cursors programs of the near-future large synoptic sky survey programs like SkyMapper, Pan-STARRS and LSST. With this goal in mind, we undertook photometric and spectroscopic follow-up observations of three recently reported unidentified transients. 
For two sources, \ntwo\ and \nthree, we show that unfortunate coincidences lead to their previous designation as transients. While the former is now interpreted as the spatial coincidence of a solar system object with  faint background star, the  latter is merely a cataclysmic variable unfortunately caught in and out of eclipse.  The third candidate, \RT\ is identified as an SU~UMa-type dwarf novae with quiescent brightness of $R\sim22.7$ and an outburst amplitude of $\sim5$\,mag. The fourth event, SDSS-SN15207, similarly shows evidence for a dwarf nova origin. Our main conclusion is that cataclysmic variables in their various avatars will contribute moderately to the population of transient objects. 

\end{abstract}

\keywords{binaries: eclipsing --- stars: dwarf novae --- novae, cataclysmic variables}

\section{Introduction}

The study of transient and variable phenomena has historically been one of the main drivers of optical astronomy. This field is now widely expected to undergo a revival, thanks to the dawn of dedicated facilities, such as SkyMapper \cite{skf05}, the Panoramic Survey Telescope and Rapid Response System \citep[Pan-STARRS;][]{k02} and the Large Synoptic Survey Telescope \citep[LSST;][]{t05}.  
A number of precursor surveys \cite[e.g.,][]{bwb04,raa05,srf05,rgs06,asg06,mab06} are already informing us of the difficulty of finding genuine new classes of transients. Solar system objects \cite{ma06} and flares from M dwarfs \cite{bwb04,kr06} are now demonstrated to be the most common variable objects. 

In this paper we report on follow-up observations of four optical transients discovered in the ROTSE-III sky patrol \cite{rag06}, the MPG/ESO Wide Field Imager (WFI) search for orphan afterglows \cite{rgs06}. and the SDSS supernovae search. Our findings show that cataclysmic variables (CVs) in general form an additional thin layer of foreground, which also must be pierced through before we can discover new classes of optical transients. 

%A few important findings have indeed been achieved by these searches \cite[c.f. the Dolichonova in M85;][]{kor06}. 

%The large  amount of false positives indicates a very  important aspect  of the existing surveys, namely to provide selection criteria which will increase the possibility to quickly distinguish the anticipated new phenomena from the perturbing foreground fog. Furthermore, the careful analysis of these foreground events will provide insightful  rate estimates, which will help to predict their contamination in future experiments.

%In a previous paper \cite{kr06} we discussed the expected confusion by flares from M-dwarfs in the Galaxy. Here, we will address another  group of events which is poised to complicate future transient searches, namely distant dwarf nova (DN) superoutbursts. In this context we report on photometric and spectroscopic follow--up observations for  three unidentified optical transients which were discovered in  the ROSTE-III sky patrol \citep[\S\ref{sec:r3}]{rag06} and the  MPG/ESO WFI search for orphan afterglows \citep[\S\ref{sec:n2} and \S\ref{sec:n3}]{rgs06}. The significance of DNe as foreground fog will be discussed in  \S\ref{sec:dis}.  

\section{\RT}
\label{sec:r3}

\subsection{Discovery}

The transient was discovered in ROTSE-III sky patrol fields on UT 2006 May 28.94 at an unfiltered magnitude of $18.2\pm0.2$ \citep{rag06}. No counterpart was found  in previous ROTSE-III observations, the Digitized Sky Survey nor the 2-Micron All-Sky Survey.   
Following the discovery, the source brightened by $\sim1.4$\,mag within a day and quickly faded back to around 18th magnitude, where it seemed to stabilize for several days. 
Observations with the 1.5-m Russian-Turkish Telescope in Bakyrlytepe, Turkey, on June 1.88 UT, 2006, detected the source at $B=18.96\pm0.02$, V=$18.31\pm0.02$ and $R_c=17.69\pm0.02$ \citep{rag06}. The   $V-R$ color of 0.6\,mag  was commented to be redder than expected for an outburst of a cataclysmic variable. Thus, the source identification remained uncertain.

\subsection{Photometry}

\begin{figure}
\includegraphics[angle=0,scale=0.6]{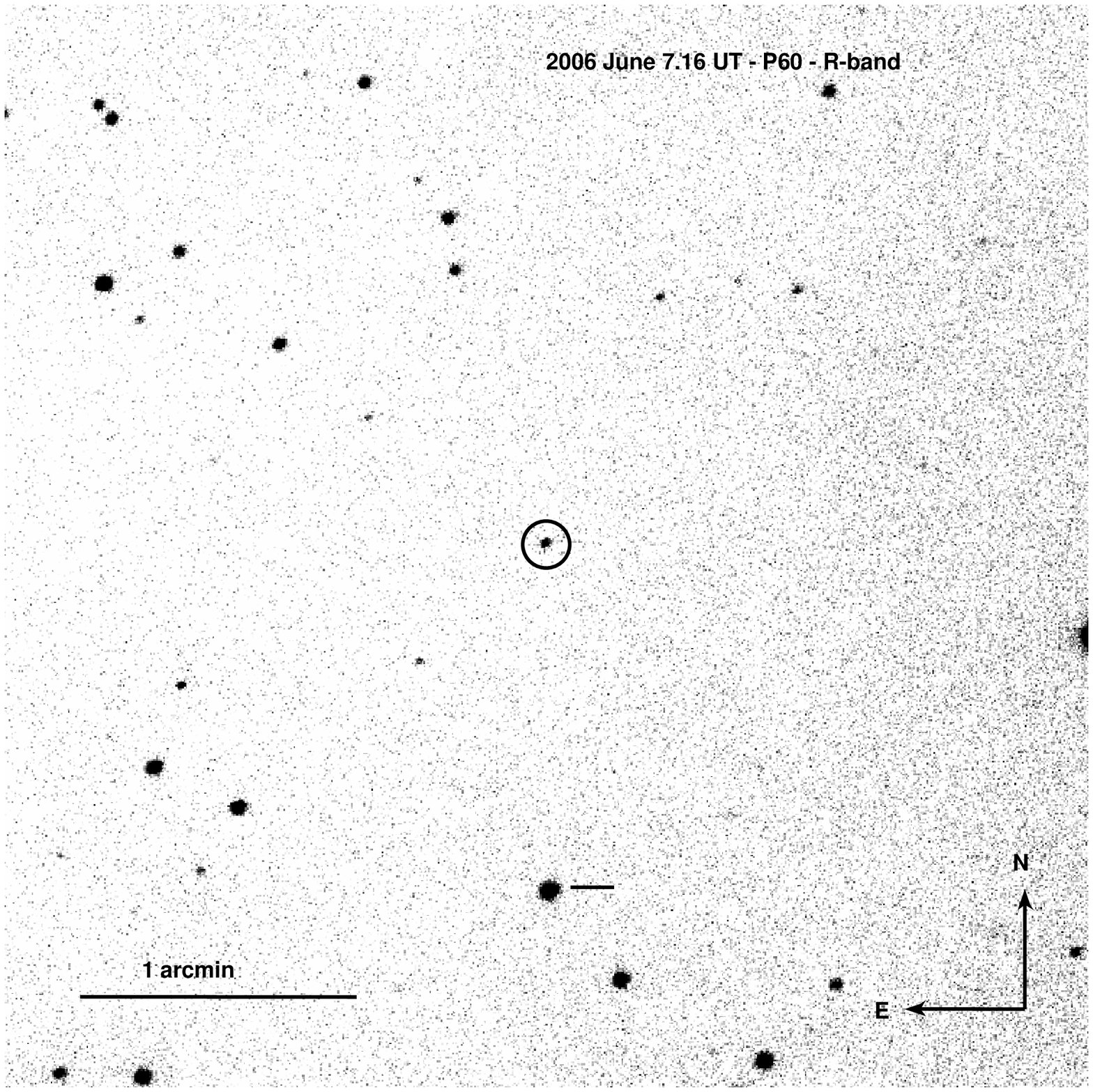}
\caption{ P60 image of the field around \RT\  taken on June 7, 2006. Astrometry was performed based on  USNO-B1.0 positions of ten stars in the neighborhood of the  transient.  The coordinates of the transient (circle) are  $\alpha_{J2000}$=16:02:13.06 \& $\delta_{J2000}$=$-$02:13:12.4. A comparison star with $R=16$\,mag located 0\farcs9 West and 75\farcs6 South  at $\alpha_{J2000}$=16:02:13:00 \& $\delta_{J2000}$=$-$02:14:28.0 is marked by the horizontal dash.
}
\label{fig:r3Fc}
\end{figure}

Our photometric follow--up of the transient (Figure~\ref{fig:r3Fc}) started on UT 2006 June 4.95, two  days after the discovery was announced and six days after the discovery itself. Our first epoch of observations were obtained with the 1-m telescope at the Wise Observatory, Israel, in the $B, V, R$ and $I$ bands. Monitoring in $g$ and $R$ with the automatic  Palomar 60-inch telescope (P60) began two days later and a final epoch was obtained with the Large Field Camera at the Palomar 200-inch telescope (P200).  A summary of the observations is presented in Table~\ref{tab:imgLog}.

\begin{figure}
\includegraphics[angle=-90,scale=0.6]{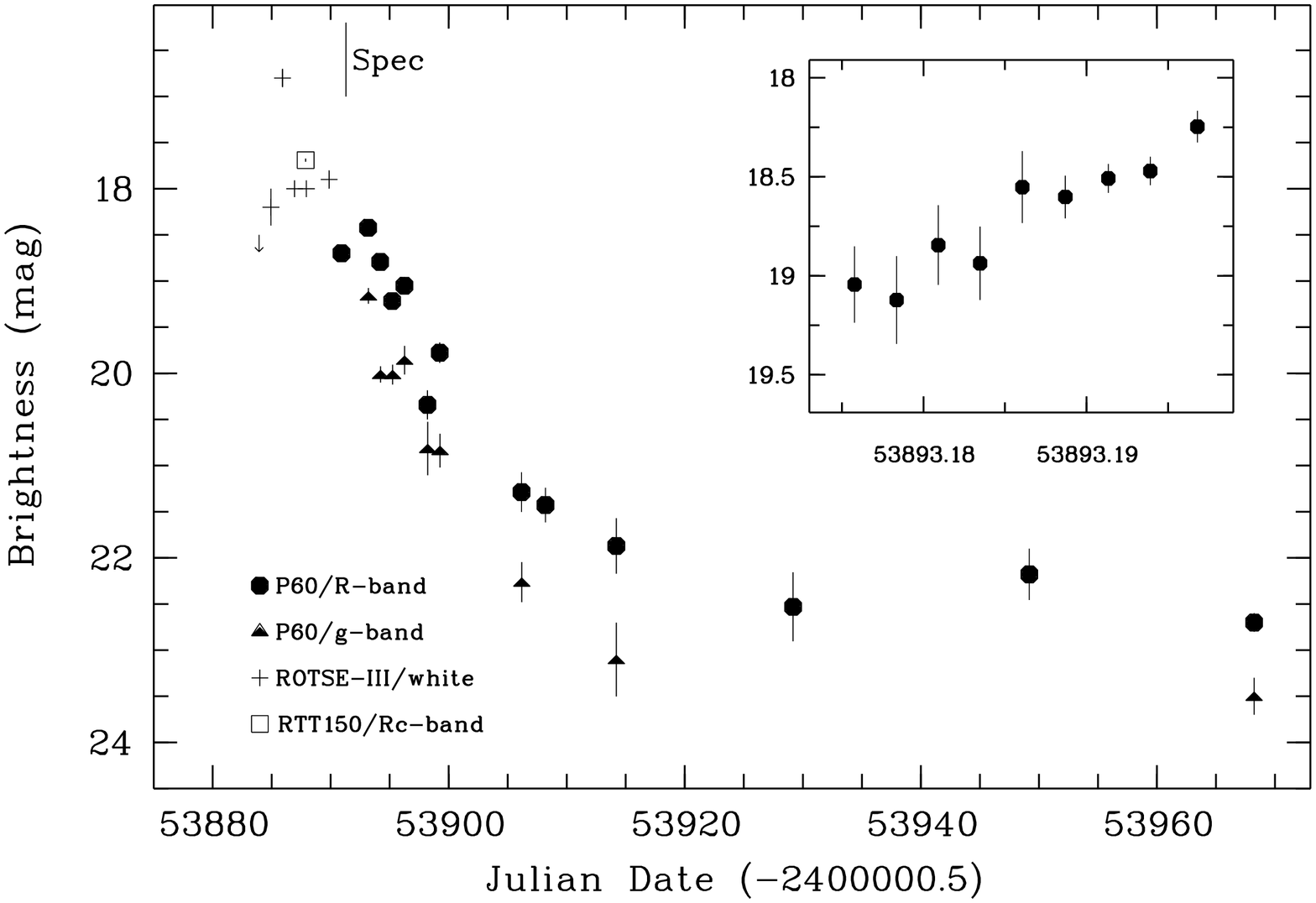}
\caption{P60 light curve of \RT\ .   Photometric calibrations (Vega system) were obtained using observations of the Landolt standard star  SA108-475 \cite{l92}.  A clear monotonic brightening by $\sim0.8$\,mag over about 30\,min was observed on June 7  (inset). Variability was also detected on June 8--10 at lower significance.   The epoch of our DBSP spectroscopy is indicated by the vertical dash.}
\label{fig:r3Lc}
\end{figure}

The first imaging epoch found the source at a brightness of $B=19.83\pm0.12$, $V=19.15\pm0.08$,  $R=18.70\pm0.05$ and $I=18.81\pm0.08$. The subsequent observations with P60 traced the decay which was interrupted by several re-brightening phases with  amplitudes of $\Delta M=0.2-0.6$\,mag and   color changes from $g-R\sim0.8$\,mag to $g-R\sim1.2$\,mag during the variations (Figure~\ref{fig:r3Lc}). The total Galactic foreground extinction in the direction of the candidate is $E(B-V)=0.276$ \cite{sfd98}. 

During the first nights of P60 observations  the brightness of the transient allowed a variability study  on short time scales using the   individual consecutive 150\,s $R$-band exposures (inset of Figure~\ref{fig:r3Lc}).

We inspected  images generated by the Near-Earth Asteroid Tracking (NEAT) program\footnote{http://skys.gsfc.nasa.gov/symorph/obs.html} for past outburst episodes at the position of the transient. No source was detected in 17 epochs between June 1998 and July 2003 with limiting magnitudes (roughly $R$-band) of 20--21.5. Similarly, no X-ray counterpart  was found in the ROSAT All-Sky survey\footnote{http://www.xray.mpe.mpg.de/cgi-bin/rosat/rosat-survey} with a limit of $6\times10^{-3}$\,count s$^{-1}$ (0.1--2.4\,keV). We conclude that the source is mostly in a quiescent state and rarely undergoes outbursts.

\subsection{Spectroscopy}

\begin{figure}
\includegraphics[angle=-90,scale=0.6]{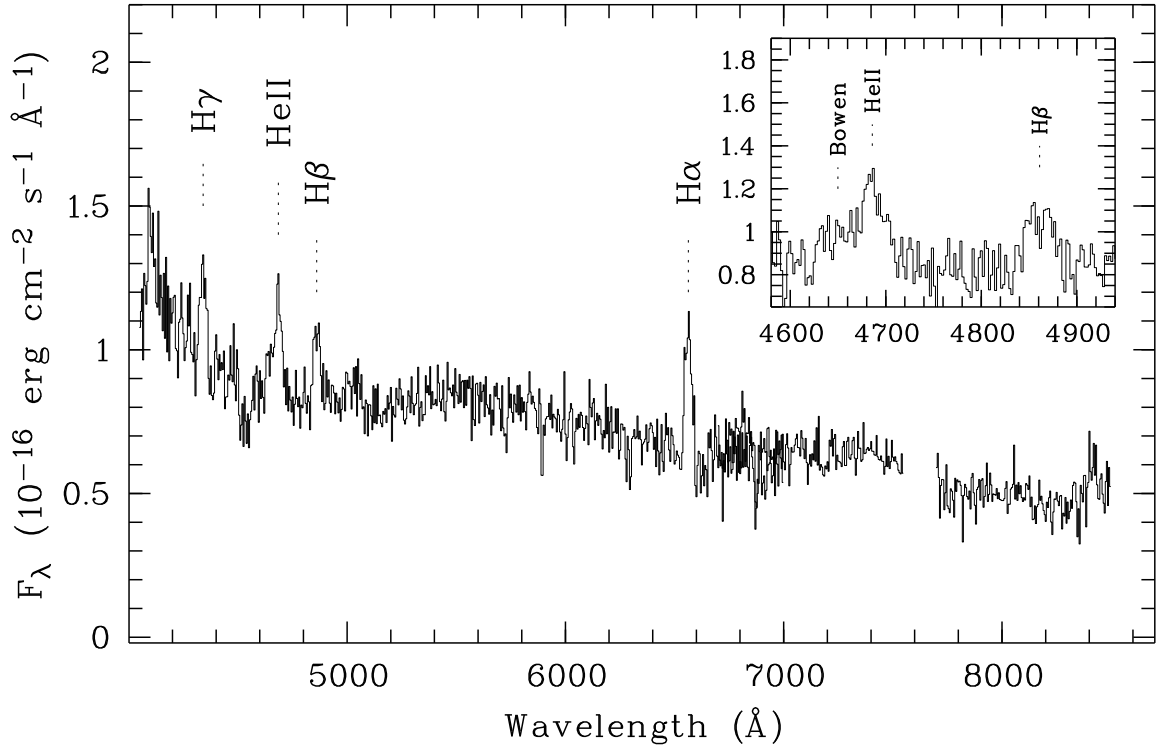}
\caption{DBSP spectrum of \RT\ . The prominent Balmer and HeII  emission features (labeled) are structured and a  double peaked shape is visible for $H\beta$ (inset) and $H\gamma$. The inset shows also the  region around the HeII line at 4686\,\AA\  revealing  a strong Bowen blend of CIII, CIV and NIII.}
\label{fig:r3Spec}
\end{figure}

Spectroscopy was obtained   with  the Double-Beam Spectrograph (DBSP) mounted on the  Hale-5-m telescope at Palomar Observatory (see Table~\ref{tab:specLog}). DBSP, as suggested by the name, operates with a dichroic and spectra can be obtained in two (blue and red) channels simultaneously. The spectrum (Figure~\ref{fig:r3Spec})  shows a flat continuum   with strong emission features which we identify with $H\alpha$, $H\beta$ and $H\gamma$. The emission lines are clearly broadened with a full width at half maximum (45$\pm$2\,\AA\ for $H\alpha$) suggesting velocities of $>$2000\,km s$^{-1}$. They are structured and double peaked  in case of H$\beta$ and H$\gamma$. In addition, a strong HeII emission line at $\lambda$4686\AA, blended with the Bowen fluorescence \cite{b34} of NIII, CIII and CIV  is detected. The line properties are summarized in Table~\ref{tab:r3Lines}.

\subsection{Identification of \RT}

The decaying light curve matches a fast classical novae (CN) burst. However, the absolute peak magnitude of a fast CN is $\sim-8$\,mag which would place the transient at a distance of 1\,Mpc. The absence of a host galaxy makes this interpretation unlikely.

However, a dwarf nova hypothesis is entirely reasonable and in agreement with the emission spectrum.
Dwarf novae constitute an important subgroup of cataclysmic variables which undergo outbursts with a brightness increase of 2--8\,mag  and repetition time scales of days  to decades. The large amplitude ($>4$\,mag, see Figure~\ref{fig:r3Lc}) of \RT\ along with the outburst duration of $>30$\,days resembles a {\it superoutburst} of SU~UMa-like DNe. Superoutburst light curves of these sources typically exhibit  a plateau phase with a slow decay of $\sim1$\,mag in 8--10\,days after the initial brightening, followed by a steep decline with periodic luminosity modulations, called {\it superhumps}.  The observed light curve differs by showing an early flare and a shallower (quasi-exponential) decay at late times.  However, we found that this temporal behavior is very similar to the superoutburst of the SU~UMa dwarf nova RU Peg in 1981 \cite{dvs85}. Although our data are too sparse to  detect clear evidence for superhumps, the re-brightenings and variability on short times scales (see inset of Fig.~\ref{fig:r3Lc}) could be interpreted as signatures of such. 

The spectrum shows strong similarities to IY UMa (an SU~UMa DN) during its superoutburst in 2000 \cite{rha05}. The large flux  ratio of HeII/$H\beta\sim1.2$ is also a popular indicator for a highly magnetic CV, e.g. a polar or an  intermediate polar. However, the temporal behavior of \RT\ with its large amplitude outburst from a long lasting quiescent state is entirely inconsistent with the long-term variability expected for  a magnetic CV. In addition, the double peaked emission lines and lack of  pronounced cyclotron lines clearly  provide further evidence that this system is a typically disk CV.

Finally, we note the large  $g-R$ color ($\sim1$) could suggest a cold accretion disk in the system resulting from  a low mass accretion rate. A low mass transfer is reminiscent of cataclysmic variables in the period gap between 2 and 3\,hours \cite[e.g.,][]{r83,rk95}, which extents to $\sim4$\,hrs for dwarf novae \cite{swc86}. Thus, we conjecture that \RT\ is a hibernating \cite{slm86} SU~UMa dwarf nova.

\section{\nthree}
\label{sec:n3}

\subsection{Discovery}

In 1999, a transient and variability survey (dedicated to the search for orphan afterglows) was performed at the MPG/ESO 2.2-m telescope equipped with the Wide Field Imager (WFI). 12\,square degrees were monitored in up to 25\,nights down to a limiting magnitude of $R=23$ \cite{rgs06}. The search netted four candidates   after filtering for moving or known solar system objects and known variables. For two sources, a classification as a CV and a flare star were suggested based on their temporal behavior.  Follow--up observations for remaining two will be described here and in the following section.

\nthree\ is an uncataloged transient found in the WFI survey when it showed a brightening from a persistent $R=19.9$ to $R=17.5$ within two days at UT 1999 June 17.14. Based on the subsequent optical decay and the detection of a faint X-ray source in the ROSAT all-sky survey, the source was preliminarily classified as a dwarf nova of the SU~UMa class.

\subsection{Photometry}

Photometry was obtained in seven epochs in May 2006 with the P60. A total of 220 $R$-band images with exposures of 150\,s were taken. The photometric calibration was tied to the  original WFI discovery image \citep[see][for details]{rgs06}. Relative photometry was applied to correct the series of P60 observations to a common reference system. The light curve revealed periodic occultations (Figure~\ref{fig:n3Lc}) and we  derived a firm ephemeris from the timing of  six individual eclipses as     

\begin{equation}\label{equ:n3}
{\rm HMJD} (T_{\rm ecl}) = 53880.2503(12) + E \times 0.099401(9).
%{\rm HMJD} (T_{\rm ecl}) = 53879.7555(12) + E \times 0.099401(9). %if RSCs was JD
%{\rm HMJD} (T_{\rm ecl}) = 53879.2555(12) + E \times 0.099401(9). %if RSCs was MJD
\end{equation}

\noindent with the numbers in brackets giving the uncertainty in the last 
digits. The accumulated cycle count between the WFI  and P60  observations is 
too large to derive  a unique period solution from 1999 to 2006. 

\subsection{Spectroscopy}

\begin{figure}
\includegraphics[angle=-90,scale=0.6]{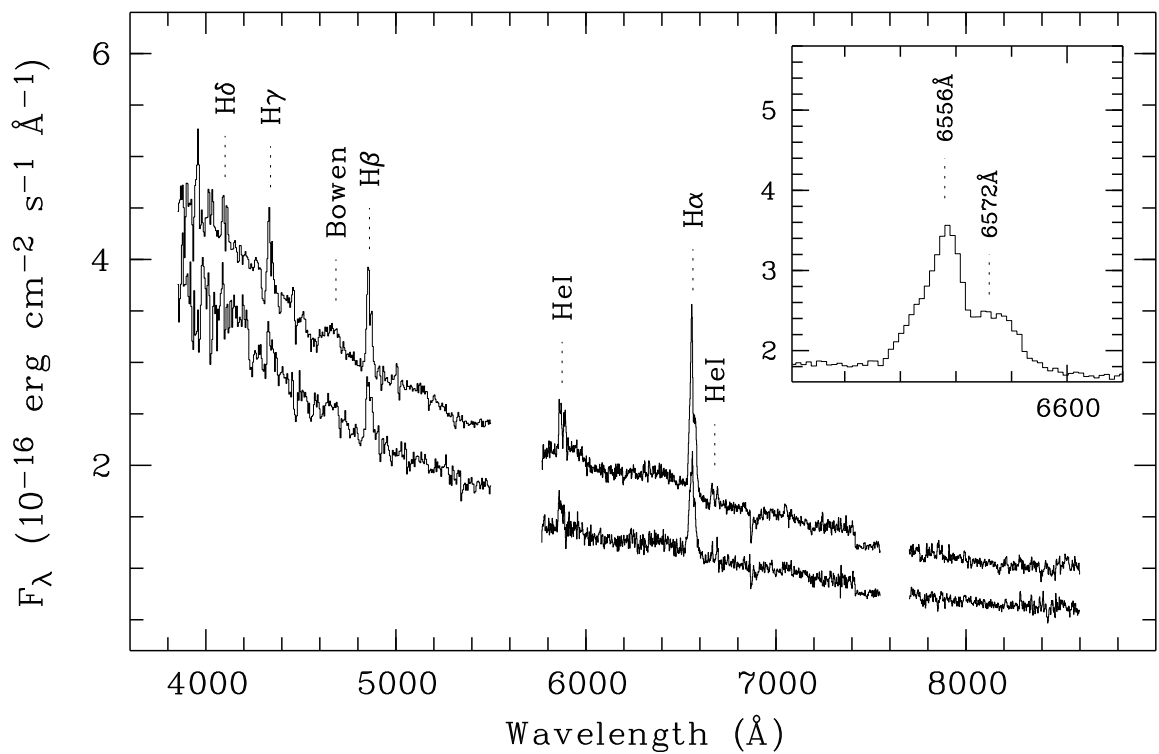}
\caption{LRIS spectra of   \nthree\ taken  2006 April 22 at orbital phases $\sim0.36$  (bottom, grey) and $\sim0.49$ (top). Prominent Balmer and HeI emission lines are indicated. The emission lines show a double peaked structure, which is shown for $H\alpha$ (phase $\sim0.49$) in the inset. In addition, a prominent Bowen blend is seen at phase $\sim0.49$  shortwards of  $H\beta$. }
\label{fig:n3Spec}
\end{figure}

Spectroscopy  was performed with the Low Resolution Imager and Spectrograph (LRIS; Oke et al. 1995) mounted on the Keck I 10-m telescope (see Table.~\ref{tab:specLog}). The LRIS light beam is split by a dichroic and spectra were obtained in the Blue and Red channel simultaneously. 
We obtained two 900\,s spectra with  mid-exposure times corresponding  to  orbital phases of $\sim0.36$ and $\sim0.49$. The spectra show blue continua with a number of strong emission features which we identify with Balmer emission ($H\alpha\dots H\gamma$) and HeI at $\lambda$5876 and $\lambda$6678. All lines are double-peaked and clearly broadened with respect to the instrumental line width (see inset of Figure.~\ref{fig:n3Spec}).  In addition, a broad emission feature centered around $\lambda=4665$\,\AA\ is found at phase 0.49. We associate this with  a Bowen blend of NIII, CIII and CIV.  A summary of the line properties is given in Table~\ref{tab:n3Lines}.

%\footnote{blue channel: $\sim$8.1\,\AA, red channel: 8.9\,\AA} 

\subsection{Identification}

\begin{figure}[t]
 \includegraphics[angle=-90,scale=0.6]{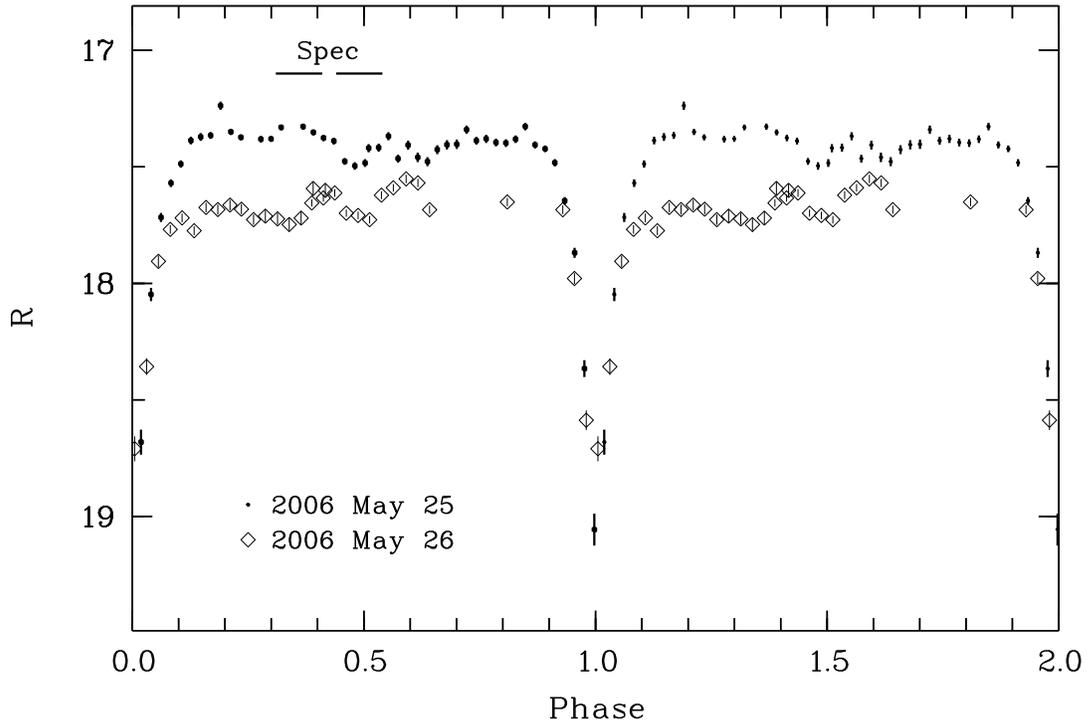}
   \caption[]{Phase-folded P60 light curves of \nthree\ using the ephemeris given in 
   Eq.~\ref{equ:n3} shown twice for clarity. Data taken on May 25th (points) and 26th (empty lozenge) reveal alterations of the out-of-eclipse variability, as well as a changes of the overall brightness level of up to 0.5\,mag.  Additional observations were performed on May 12, 13, 15, 16, \& 17, 2006 (see electronic supplement). Horizontal dashes mark the phases of the Keck/LRIS spectroscopy observations (see Figure~\ref{fig:n3Spec})}
  \label{fig:n3Lc}
\end{figure}

The periodic light curve and the spectrum  resemble those of  a dwarf nova in quiescence.  The detection of deep orbital eclipses (Figure~\ref{fig:n3Lc}), together with the double peaked emission lines  indicate a high inclination of the system. Out of  eclipse, the light curve appears flatter than typically observed in eclipsing dwarf novae. In addition, no strong asymmetry is seen 
between ingress and egress, suggesting that the hot spot, the region where the gas stream collides with the accretion disk, would be weaker than usual. This is in agreement with the low mass transfer rate assumed for CVs in the period gap, in which \nthree\ falls.  

The double peaked emission lines (Figure~\ref{fig:n3Spec}), the  absence of pronounced cyclotron features and the non-detection of  HeII indicate a disk system. The detection of the Bowen blend around $\lambda$4665 bears evidence of irradiation in the disk or the secondary. HeII emission, which leads to the Bowen fluorescence, is not detected, though. This suggests a considerable HeII opacity.

 The orbital period of $\sim143$\,min  implies that observations taken at the same UT but over  different nights,  will result in the source being observed at  nearly identical orbital phases. In retrospect, we now realize that the original survey data were obtained at phases at which the object exhibited extreme brightness. 
 The  maximum brightness in the WFI detections of $R=17.5$ corresponds to the median brightness out of eclipse (see Figure.~\ref{fig:n3Lc}) and the (falsely interpreted) quiescent brightness of $R=19.9$ is similar to what is observed during the eclipse. 

Summarizing, we found no evidence for a dwarf nova outburst of \nthree\ and thus are left with the rough classification of the source as a disk CV.  Indeed,  the brighter quiescent magnitude than previously thought, implies  $L_X/L_{opt}\sim0.1$, consistent with a larger variety of CVs.

\section{\ntwo}
\label{sec:n2}

\subsection{Discovery}

The source was detected on June 26.07 UT, 1999, at a maximum brightness of $R=19.9$. As a result of the scheduling, no observations  could be obtained during a subsequent decay. However, a faint uncataloged and persistent object ($R=21.3$) was found, offset by 0\farcs8 from the location of the transient source.  This apparent association allowed for a possible extragalactic origin of the transient with the persistent source being the underlying host galaxy. It was therefore interpreted as the best candidate for an orphan afterglow found in the survey. However, given the sparse data coverage, other interpretations (extra-galactic, galactic and even solar system) could not be excluded.

\subsection{Spectroscopy}

\begin{figure}
\includegraphics[angle=-90,scale=0.6]{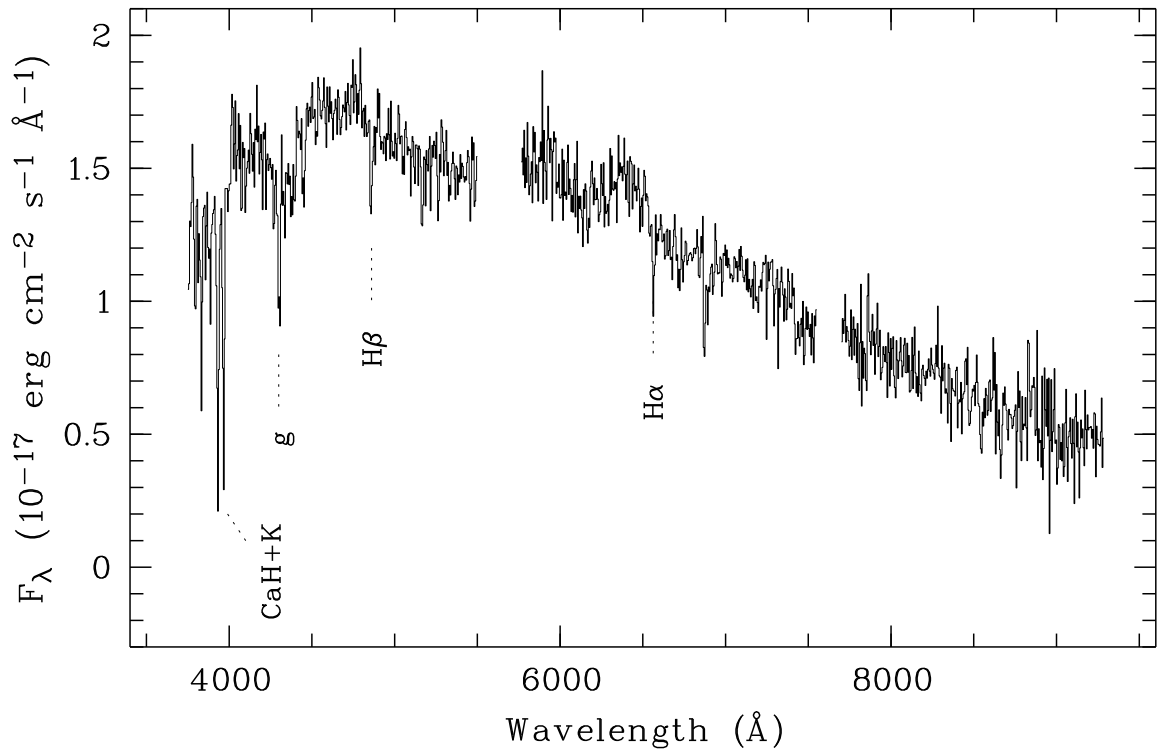}
\caption{LRIS spectrum of the quiescent counterpart to \ntwo\ taken 2006 April 23. Prominent absorption features are indicated. }
\label{fig:n2Spec}
\end{figure}

Spectroscopy of the of the persistent source associated with \ntwo\  was performed with LRIS.
We find strong absorption features which we identify with Ca H\&K, g-band, $H\beta$ and $H\alpha$ (Fig.~\ref{fig:n2Spec}). The narrow lines allow to estimate the velocity with respect to the local standard rest frame to 0$\pm$50\,km s$^{-1}$. No significant emission lines are detected.

\subsection{Identification}

The spectrum is consistent with that of a galactic G-star. Thus, we can rule out an extra-galactic origin for the persistent source. As a result, there is little justification to assume that the transient source was extra-galactic. Instead the detection was likely the misleading alliance of a moving solar system object  superimposed on a background star.  The position in the ecliptic plane (ecliptic coordinates $\lambda=208.5\,^\circ$ \& $\beta = -11.6\,^\circ$) supports this scenario. 

\section{SDSS-SN15207}

\subsection{Discovery}
SDSS-SN15207 was discovered on 2006 October 11 as part of the Sloan
Digital Sky Survey (SDSS) search for supernovae at RA(J2000)=20$^h$58$^m$43$^s$45, Dec(J2000)=01$^\circ$02$^\prime$47$^{\prime\prime}$0. SDSS imaging prior to the
eruption shows no underlying host galaxy to a limit of $r$=24. The SDSS optical light curve\footnote{http://sdssdp47.fnal.gov/sdsssn/candidates/examineCand.php?cid=15207} showed a peak magnitude of $r$=19.8 on Oct 12 followed by a decline by 1.2\,mag over 5\,days. Continued monitoring with P60 during the first 80\,days did not detect additional outbursts to a limiting magnitude of $r$=22.

\subsection{Spectroscopy}
\begin{figure}
\includegraphics[angle=-90,scale=0.6]{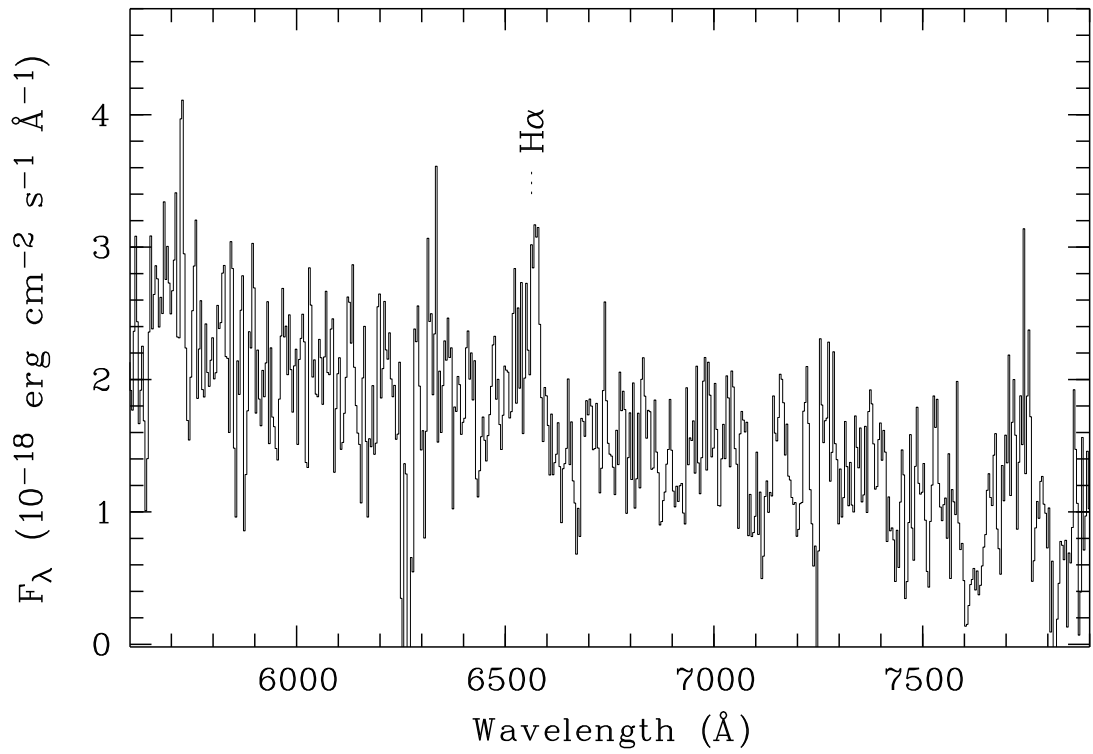}
\caption{Gemini GMOS spectrum of SDSS-SN15207 obtained 2006 November 17, 36\,days after the discovery.}
\label{fig:15207}
\end{figure}

We obtained a spectrum with the Gemini Multi Object Spectrograph (GMOS) on Gemini-South on 2006 November 17 (Table~\ref{tab:specLog}). At this time, the event had faded to $r\sim23$. This resulting low-signal-to-noise  spectrum  (Figure~\ref{fig:15207}) shows a blue continuum and a single emission line (FWHM$=1000\pm200$\,km/s), which we identify with H$\alpha$.  The line centroid suggest a Galactic origin.

We also obtained a 4.77\,ks exposure with the
X-ray telescope aboard Swift on 2006 November 4. Not  counterpart was found with a 2$\sigma$ flux limit of $>6\times10^{-14}$\,ergs cm$^{-2}$ s$^{-1}$ (0.2--10\,keV). 

\subsection{Identification}

The peak magnitude of this transient rules out the possibility that it is a  nova in the Milky Way. The optical light curve, in particular, the steep decline of 1.2 mag in 5 days and 1.6 mag in 9 days does not suggest a typical supernova either. On the other hand, the observations are consistent with a super outburst ($>4$\,mag) of a Galactic dwarf nova. The non-detection  of the quiescent counterpart to $r=24$ and the lack of significant foreground extinction in the optical colors and spectrum, suggest SDSS-SB15207 to have be located a a relative large distance from the earth. Assuming a absolute quiescence magnitude of $R=12$ \citep[white dwarf with $T=11000$\,K, $M=0.6$\,\msun;][]{bwb95}, the lower limit on the distance is 2.5\,kpc.

%At 2.5\,kpc this
%corresponds to a limit on the X-ray luminosity of $>4\times10^{31}$ erg s$^{-1}$, is consistent with typical dwarf novae \cite{pcm05}.

\section{Discussion and Conclusion}
\label{sec:dis}

Here we reported on photometric and spectroscopic follow--up observations for four recently communicated transient sources. We found that \ntwo, the best orphan afterglow candidate detected in the WFI survey \cite{rgs06}, was with high probability a misidentified solar system object spatially coincident with a background source. Next, the potential outburst light curve of a second source from the same survey, \nthree,  was recognized as serendipitous observations in and out of  eclipse of a high-inclination cataclysmic variable. Finally, \RT, discovered by ROTSE-III \cite{rag06} and (potentially SDSS-SN15207) were identified as the superoutbursts of  SU~UMa-like dwarf nova.

In the following, we restrain from  exploring the two WFI sources any further, except noting again the interesting coincidences  which had lured us to incorrect conclusions previously. Instead, we  want to address the implications of the faint  quiescent magnitudes of \RT\ ($g\sim23$) and SDSS-SN15207 ($r>24$). We start by noting that a very similar event (1955+22C VAR VUL 05 \footnote{AAVSO Alert Notice 325}) has been discovered in August 2005 with  15.8\,mag at maximum and an amplitude of at least 9\,mag.  

This quiescent faintness of these three events indicates an important factor for future wide-field transient searches, namely, dwarf novae which stay most of the time below the limiting  magnitude of a survey, but  appear as new transient sources during outbursts.  This includes CVs which appear faint in quiescence because of  large distance or Galactic extinction and  intrinsically low-magnitude systems. The latter group is dominated by systems with low mass transfer, typical for CVs in the period gap and in hibernation \cite{slm86}. As a matter of fact, the vast majority of non-magnetic CVs with $P_{orb}<3$\,hrs are dwarf novae ($\sim83$\,\%) while this is true only for $\sim36$\,\% with $P_{orb}>3$\,hrs \cite{agr06}.

As an example, let us consider   the proposed 3$\pi$ steradian survey with PanSTARRS-1 (PS1) which is expected to reach a limiting magnitude of $R\sim23.2$  and $R=24.9$ in a single 30\,s  and combined images over 3\,yrs, respectively. In such a survey, sources similar to \RT\ at twice the distance  would no longer be detectable in quiescence, but only during the rare outburst periods. This directly leads to the question of how important these events will be in future wide-field surveys. 

The expected number of DNe events in an all-sky snapshot  can be obtained rather crudely, given the large number of involved parameters (e.g., intrinsic quiescent magnitude, superoutburst amplitude, superoutburst cycle).  Let the  quiescent absolute magnitude of the DN be  M$_R\sim$12 \citep{bwb95}. For the volume distribution we assume a local scale height of 300\,pc perpendicular to the Galactic disk  and isotropy in the radial direction in the disk. Furthermore, we consider for simplicity an upper limit for the distance in the Galactic Plane of 2\,kpc, above which extinction will dominate at low Galactic latitude. For the observed local density\footnote{Note that the observationally derived space densities are uncertain by at least a factor of ten in both directions. A detailed assessment of possible selection effects in surveys can be found in Patterson 1984.} of dwarf novae we use $\rho\sim3\times10^{-5}$\,pc$^{-3}$ \cite{sbb02} and theory allows for even higher values \cite[$10^{-4}$ to $10^{-3}$\,pc$^{-3}$;][]{k93,dkr93}.  The resulting  number of expected dwarf novae in the considered volume, independent of their activity, is given in Table~\ref{tab:DNstats} for three different ranges of Galactic latitude.

How many of those will be in superoutburst at any given time? Assuming a mean cycle duration of 1\,yr and a  mean plateau length of 10\,days we find that between 40  and 1.5$\times10^3$ events, depending on the Galactic latitude and assumed space density, can be expected in an all-sky snapshot with maximum depth. Most of these ($\sim86\,\%$) will be at low and medium Galactic latitude and only a small fraction ($\sim14\,\%$) is expected to occur at $|b|>45^\circ$.  

\begin{figure}
\includegraphics[angle=-90,scale=0.5]{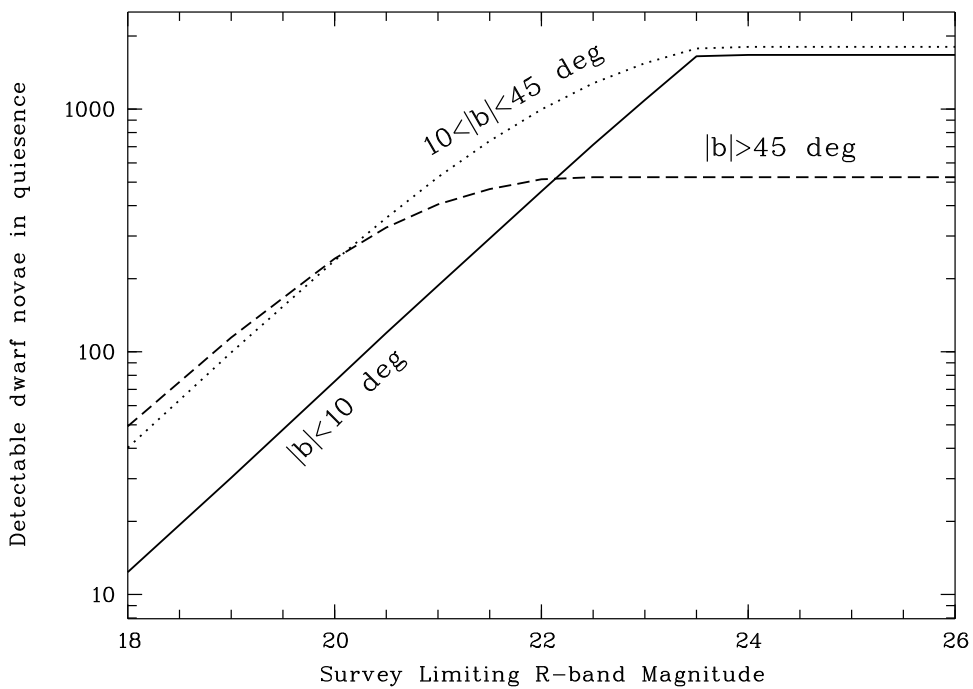}
\caption{Predictions for the number of dwarf novae which can be detected in a survey with a given limiting $R$-band magnitude. The solid, dashed and dotted lines show the expectations for three ranges of Galactic latitude. Note, that the sharp turnover for $|b|<10^\circ$ is caused by the distance limit in the Galactic Plane (2\,kpc) which was assumed to account for foreground extinction. At $|b|>45^\circ$ nearly all quiescent system will be detected above $R=22$.}
\label{fig:prediction}
\end{figure}

The space density of CVs decreases rapidly with distance from the Galactic Plane. Similarly, at low Galactic latitude extinction limits the observable. As a result, the number of dwarf novae with quiescent brightness below the survey limiting magnitude (here  $R\sim25$) will be negligible. However, shallower surveys will detect less quiescent systems as indicated in Figure~\ref{fig:prediction}. Thus, mainly these surveys will be contaminated by dwarf novae superoutbursts with undetected quiescent counterparts.

The yearly all-sky rate of less than $4\times10^4$ detectable superoutbursts  is in the range of  the local volume rates (Gpc$^{-3}$ yr$^{-1}$) for core-collapse SNe \citep[$\sim5\times10^4$;][]{cet99} but considerably lower than the rates of  novae \citep[$\sim10^8$;][]{kr06} and Galactic M-dwarf flares \citep[$10^8$\,yr$^{-1}$;][]{kr06}. In addition, one has to remember that the theoretical CV space densities typically imply a large fraction of systems with cycle periods of years to decades (e.g., WZ Sge). Thus, the yearly all sky rate is probably much less than value given above. This suggests that dwarf nova superoutbursts will not produce a dominant  contamination in future wide-field surveys. However, they will still be of concern for observers mining the optical transient sky.

How can the small number of dwarf novae outbursts   be distinguished in practice from other kinds of interesting transients?  The most important factors will be the choice of cadence and the field selection. 
A good sign for a DN will be a location close to the Galactic disk or inner Galaxy. Indeed, the positions of \RT\ ($l=8\,^\circ$, $b=35\,^\circ$) and also of 1955+22C VAR VUL 05 ($l=60\,^\circ$, $b=-4\,^\circ$) are consistent with being part of the disk population. However, the Galactic Plane is known to harbor plentiful of variable and transient sources. Choosing the right observing strategy (cadence, filter) will help to distinguish  between short lived phenomena (e.g., flare stars) and dwarf novae. The main confusion at high Galactic latitude will be by novae and supernovae. Novae, however, will be distinguishable by their enormous brightness (if Galactic) or by their association with a nearby galaxy. Supernovae may similarly identified by their host galaxies. Spectroscopy of the remaining candidates  would require significant amounts of time on large telescopes and is thus not a feasible approach. Photometric colors vary considerably from one event to another and will therefore not be useable as reliable indicators either. The best confirmation, however, can be obtained by photometric follow-up searches for the expected superhump structures. 

We close this paper noting the role of future large surveys for our general understanding of cataclysmic variables. Complete wide-area surveys on long times scales will allow the in-depth study of  the distribution of orbital periods and outburst behaviors (amplitudes, cycles). This will provide important informations on the evolution of CVs and the processes of mass accretion in these systems.

%As also discussed elsewhere \cite{kr06}, surveys for new optical transient phenomena will require specific strategies to offer maximum success. Large Area surveys with transient search as secondary science will be drowning in false positives, though. The best line of defense for avoiding the contribution of dwarf nova outbursts (as well as flares from M dwarfs and other Galactic contaminants) will be a wise choice of directions, preferentially a high galactic latitude ($|b|>30$). Even so, a small  rate of DNe   might also occur in the Galactic halo. Repeating observations of target fields will help to identify sources with short activity cycles by  detecting recurrent superoutbursts. However the superoutburst cycle lengths of up to decades will prevent the identification of all such events.

%\noindent{\it Note added after submission:}  A dwarf nova candidate was recently found as a  new high-amplitude ($\Delta m\sim7$\,mag)  transient in the Catalina Sky Survey (Christensen et al., CBET \#746). We obtained subsequent  spectroscopy at P200 and reported a B3-4V stellar like spectrum 23\,days after the discovery (Rau et al., ATel \#951)  which is consistent with a dwarf nova during outburst (Ayani  et al., CBET \#753). Later spectroscopy revealed emerging H$\alpha$ emission and supported this identification.

\acknowledgments

We acknowledge helpful discussions with J. Greiner, E. Mason and L. Bildsten. We thank M. Sako and A. Becker for the timely alert about the eruption of  SDSS-SN15207. Furthermore, we thank the anonymous referee for constructive comments. This work is supported in part by grants from the National Science Foundation and NASA.

\clearpage

% ---------------------------------------------------------------------------------------------------------
% ---------------------------------------------------------------------------------------------------------

% ---------------------------------------------------------------------------------------------------------
% ---------------------------------------------------------------------------------------------------------

% ---------------------------------------------------------------------------------------------------------
% ---------------------------------------------------------------------------------------------------------

% ---------------------------------------------------------------------------------------------------------
% ---------------------------------------------------------------------------------------------------------

% ---------------------------------------------------------------------------------------------------------
% ---------------------------------------------------------------------------------------------------------

% ---------------------------------------------------------------------------------------------------------
% ---------------------------------------------------------------------------------------------------------

% ---------------------------------------------------------------------------------------------------------
% ---------------------------------------------------------------------------------------------------------

% ---------------------------------------------------------------------------------------------------------
% ---------------------------------------------------------------------------------------------------------

% ---------------------------------------------------------------------------------------------------------
\begin{table}
\begin{center}
\caption{\RT\  imaging log (for electronic version).\label{tab:imgLog}}
\begin{tabular}{ccccc}
\tableline\tableline
MJD & Telescope & Filter & Exposure (s) & Brightness (mag)\\
\tableline
53890.94 & Wise & B & 840 & 19.83$\pm$0.12\\
53890.94 & Wise & V & 600 & 19.15$\pm$0.08\\
53890.93 & Wise & R & 600 & 18.70$\pm$0.05\\
53890.96 & Wise & I  & 600 & 18.81$\pm$0.08\\
53893.19 & P60 & g & 1350 & 19.16$\pm$0.08\\
53893.19 & P60 & R & 1350 & 18.42$\pm$0.05\\
53894.21 & P60 & g & 1350 & 20.01$\pm$0.09\\
53894.21 & P60 & R & 1350 & 18.79$\pm$0.05\\
53895.22 & P60 & g & 1350 & 20.01$\pm$0.11\\
53895.22 & P60 & R & 1350 & 19.22$\pm$0.07\\
53896.26 & P60 & g & 1350 & 19.86$\pm$0.15\\
53896.26 & P60 & R & 1350 & 19.05$\pm$0.09\\
53898.21 & P60 & g & 1350 & 20.82$\pm$0.29\\
53898.21 & P60 & R & 1350 & 20.34$\pm$0.16\\
53899.25 & P60 & g & 1350 & 20.84$\pm$0.22\\
53899.25 & P60 & R & 1350 & 19.78$\pm$0.11\\
53906.16 & P60 & g & 1350 & 22.26$\pm$0.21\\
53906.16 & P60 & R & 1350 & 21.29$\pm$0.22\\
53908.20 & P60 & R & 1350 & 21.43$\pm$0.19\\
53914.29 & P60 & g & 1350 & 23.10$\pm$0.40\\
53914.29 & P60 & R & 1200 & 21.87$\pm$0.29\\
53929.19 & P60 & R & 2700 & 22.53$\pm$0.37\\
53949.19 & P60 & R & 2700 & 22.18$\pm$0.28\\
53968.23	& P200 & g' & 600 & 23.5$\pm$0.2\\
53968.24	& P200 & r' &  600 & 22.7$\pm$0.1\\
\tableline
\end{tabular}
\tablecomments{All magnitudes are given in the Vega sytem.}
\end{center}
\end{table}
% ---------------------------------------------------------------------------------------------------------

%----------------------------------------------------------------------
\begin{table}
\begin{center}
\caption{Log of the spectroscopic observations.}
\begin{tabular}{lcccc}
\hline 
                    & Instrument & Setup        & Exposure  (s)   \\
\hline\hline
{\bf \RT:}\\
2006 June 5 & Hale-5m/DBSP & 300/3990 & 2100 \\
		     & Hale-5m/DBSP & 316/7500 & 2100 \\
\hline
{\bf \nthree:}\\
2006 April 22 &  Keck/LRIS & 400/3400 & 2$\times$900   \\
                          & Keck/LRIS & 400/8500 (7608\,\AA) & 2$\times$900   \\
{\bf \ntwo:}\\
2006 April 23  & Keck/LRIS & 400/3400 & 3600   \\
                          &  Keck/LRIS &  400/8500 (7608\,\AA) & 3600  \\
\hline
{\bf SDSS-SN15207:}\\
2006 November 17 & Gemini-S/GMOS & 150/7170 & 2$\times$1750\\                         
\hline 
\hline
\end{tabular}
\tablecomments{  In column three we  list  the grism/grating (number of  grooves/blaze angle) and (for the red DBSP and LRIS gratings)  the central  wavelength.}
\end{center}
\label{tab:specLog}
\end{table}
%----------------------------------------------------------------------

%----------------------------------------------------------------------
\begin{table}
\begin{center}
\caption{\RT\ emission line characteristics.\label{tab:r3Lines}}
\begin{tabular}{lccc}
\tableline\tableline
Species &  Flux & --EW & FWHM \\
                  & $\times10^{-15}$\,erg/cm$^2$/s  & (\AA) & (\AA)  \\
\tableline
H$\gamma$ $\lambda$4340  & 1.6$\pm$0.3 & 19$\pm$4 & 42$\pm$7\\
%NIII $\lambda$4637 & 4635 & 3.7$\pm$ & -9$\pm$ & 21$\pm$ & 1360$\pm$\\
%CIII $\lambda$4650 & & $\pm$ & $\pm$ & $\pm$ & $\pm$ \\
%CIV  $\lambda$4658 & 4655 & 2.3$\pm$ & -5$\pm$ & 19$\pm$ &1225$\pm$\\
HeII $\lambda$4686  & 1.9$\pm$0.3 & 25$\pm$5 & 42$\pm$5 \\
H$\beta$ $\lambda$4861  & 1.7$\pm$0.2 & 22$\pm$5 & 41$\pm$7 \\
H$\alpha$ $\lambda$6563  & 2.6$\pm$0.2 & 58$\pm$5 & 45$\pm$2 \\
\tableline
\end{tabular}
\end{center}
\end{table}
%----------------------------------------------------------------------

%----------------------------------------------------------------------
\begin{table}
\begin{center}
\caption{\nthree\ emission line characteristics.\label{tab:n3Lines}}
\begin{tabular}{lcccc}
\tableline\tableline
& \multicolumn{2}{c}{Phase $\sim0.36$} & \multicolumn{2}{c}{Phase $\sim0.49$}\\
Species & Flux & --EW & Flux & -EW\\
                   & $\times10^{-15}$\,erg/cm$^2$/s  &  (\AA) & $\times10^{-15}$\,erg/cm$^2$/s & (\AA)\\
\tableline
H$\delta$ $\lambda$4102  & $<$1.2 & $<$5 & 1.4$\pm$0.2 & -3$\pm$1\\
H$\gamma$ $\lambda$4340  & 1.5$\pm$0.5 & 6$\pm$2 & 2.7$\pm$0.2 & 4$\pm$1 \\
Bowen $\lambda$4665 & $<$2.2 &$<$9 & 5.5$\pm$0.5 & 15$\pm$5 \\
 H$\beta$ $\lambda$4861 &  2.8$\pm$0.3 & 13$\pm$1 & 3.7$\pm$0.2 & 13$\pm$1\\
HeI $\lambda$5876  & 1.0$\pm$0.2 & 7$\pm$1 & 1.5$\pm$0.4 & 6$\pm$1 \\
H$\alpha$ $\lambda$6563  & 3.9$\pm0.3$ & 34$\pm$3 & 5.4$\pm$0.3 & 34$\pm$3\\
HeI $\lambda$6678  & 0.4$\pm$0.1 & 3.4$\pm$0.4 & 0.6$\pm$0.1 & 3.6$\pm$0.5\\
\tableline
\end{tabular}
\end{center}
\end{table}
%----------------------------------------------------------------------

%----------------------------------------------------------------------
\begin{table}
\begin{center}
\caption{Dwarf novae number counts and outburst rates.\label{tab:DNstats}}
\begin{tabular}{ccccc}
\tableline\tableline
Galactic latitude & \multicolumn{2}{c}{$N_{DN}$$^1$} & \multicolumn{2}{c}{$N_{SO}$$^2$}\\
& $\rho=3\times10^{-5}$\,pc$^{-3}$ & $\rho=10^{-3}$\,pc$^{-3}$ & $\rho=3\times10^{-5}$\,pc$^{-3}$ & $\rho=10^{-3}$\,pc$^{-3}$ \\
\tableline
$|b|<10^\circ$ & $\sim1.7\times10^3$ & $\sim5\times10^4$ & $\sim40$ & $\sim1.5\times10^3$\\
$10^\circ<|b|<45^\circ$ & $\sim1.8\times10^3$ & $\sim5\times10^4$ & $\sim40$ & $\sim1.5\times10^3$\\
$|b|>45^\circ$ & $\sim5\times10^2$ & $\sim1.5\times10^4$ & $\sim15$ & $\sim4\times10^2$ \\
\tableline
\end{tabular}
\tablecomments{$^1$: total number of dwarf novae within the considered volume. See text for details. $^2$: number of superoutbursts in an all-sky snap shot. Note, that the theoretical DN density includes a large population of events with superoutburst cycles of decades. Thus, the presented numbers are optimistic. Results will change if a different scale height \cite[e.g., 150\,pc;][]{p84} is adopted.}
\end{center}
\end{table}
%----------------------------------------------------------------------

\end{document}